# Active compensation of extrinsic polarization errors using adaptive optics

YUANYUAN DAI[1], CHAO HE[1], JINGYU WANG[1], RAPHAËL TURCOTTE[1,2], LEWIS FISH[1], MATTHEW WINCOTT[1], QI HU[1], AND MARTIN J. BOOTH[1,*]

[1]*Department of Engineering Science, University of Oxford, Parks Road, Oxford, OX1 3PJ, United Kingdom*
[2]*Department of Pharmacology, University of Oxford, Mansfield Road, Oxford, OX1 3QT, United Kingdom*
*\* martin.booth@eng.ox.ac.uk*

**Abstract:** We present a scheme for active compensation of complex extrinsic polarization perturbations introduced into an optical system. Imaging polarimeter is used to measure the polarization state across a beam profile and a liquid crystal spatial light modulator controls the polarization of the input beam. A sequence of measurements permits determination of the birefringence properties of a perturbing specimen. The necessary correction is calculated and fed back to the polarization modulator to compensate for the polarization perturbation. The system capabilities are demonstrated on a range of birefringent specimens.

## 1. Introduction

Adaptive optics has predominantly focused on the correction of phase aberrations [1-3], but there are many optical systems where polarization errors can also have detrimental effects on performance [4, 5]. While polarization control of a whole beam is a commonplace component of many optical systems, spatially variant control of polarization across a beam profile is also used, such as for manipulation of vector beams [6-9]. However, there are many reasons why spatially varying polarization errors can be introduced in a system, such as the effects of internal stresses in optical components, dielectric coatings, or birefringent materials [10, 11]. In microscopes, for example, beam profile dependent polarization errors can arise from specimen birefringence or Fresnel effects when focusing at high numerical aperture through interfaces between the coverglass and immersion or mounting media [12-16]. It is known that in certain microscope applications, polarization plays an important role. For example, in STED microscopy, polarization errors can cause a non-zero intensity at the center of the depletion beam [17]; in structured illumination microscopy, the contrast of the sinusoidal illumination patterns is dependent upon the use of appropriate polarization [18, 19]. It is clear therefore that the adaptive control of polarization could be beneficial in such cases.

Conventional adaptive optics involves the measurement and control of phase aberrations in the pupil of the system, which can be detected using a wavefront sensor and corrected using an adaptive element, such as a deformable mirror or a spatial light modulator (SLM). The equivalent concept for polarization would involve an imaging polarimeter and a device for adaptive polarization control. A wide range of imaging polarimeters could be employed for this purpose by using prisms [20, 21], liquid crystal variable retarders [22] or interferometric methods [23-25]. Polarization control has been implemented using liquid crystal SLMs in various configuration for different applications such as investigating the principal modes in a multimode fiber [26], arbitrary polarization generation [27, 28] and quantum logic [29]. Some polarization rectifiers have been implemented to correct the polarization state due to static errors introduced by microscope lenses or in other complex optical systems [30-33]. Mostly beam control has been implemented in open loop manipulation of the polarization state. Closed loop control has also been used for compensation of intrinsic errors in beam generation systems [33]. However, the potential exists for feedback correction using a combination of polarimeter and modulation to compensate for externally induced polarization disturbances.

In this paper, we introduce an adaptive polarization control (APC) system, encompassing both hardware and algorithms, which allow us to control the output polarization profile of a beam, even when perturbed by a specimen. The APC system consists of two hardware modules: a polarization state generator (PSG), in the form of a dual pass liquid crystal SLM, and a polarization state analyzer (PSA), comprising a rotating waveplate based imaging Stokes polarimeter. The APC system can extract the birefringence profile of a specimen and then adaptively manipulate the output polarization state through control of the input beam using the PSG.

The operation is described in terms of Jones calculus. As the Jones matrix of an unknown birefringence cannot be determined unambiguously through a single polarimeter measurement in the APC system, we investigate different strategies that can be employed depending upon the amount of prior information one has about the specimen. Through the assumption that the specimen is purely birefringent and introduces no diattenuation, we find that between one and three PSA measurements are required, dependent upon the knowledge of the specimen. We demonstrate four kinds of samples for system validation: a vortex half wave retarder, a stressed plastic plate, a customized liquid crystal device and excised mouse brain tissue. In each case we show how the polarized component of the output light can be restored through feedback correction to the desired state, despite the perturbation.

## 2. Methodology

The schematic basis of the system is shown in Fig. 1. The purpose of the system is to generate a desired polarization state at the output, even when an unknown birefringent perturbation is introduced by an object. Compensation for the perturbation is implemented by producing the necessary polarization state using the PSG at the input. A sequence of measurements is taken, the number of which depends on the degree of prior information that is known about the perturbation. Using these measurements, the polarizing characteristics of the object are determined and the necessary input state is generated.

### 2.1 Theoretical model

We based the theoretical model on the scheme in Fig. 1. According to Jones calculus, the polarization of the output light is represented as Eq. (1), assuming there is no loss and attenuation during beam propagation:

$$\boldsymbol{J}_{out} = \boldsymbol{J}_{sample}\boldsymbol{J}_{mod} \tag{1}$$

where $\boldsymbol{J}_{out}$ is the normalized Jones vector of the output light. $\boldsymbol{J}_{sample}$ is the Jones matrix of the birefringent perturbation and $\boldsymbol{J}_{mod}$ is the Jones vector of the modulated input light, which is created by the PSG and which we will discuss in next section. In all cases presented here, the matrices are functions of the spatial coordinates across the beam profile. For brevity, we omit the explicit dependence on these coordinates from the notation. For the normalized Jones vector, the absolute phase is removed and retardance information is kept, which is sufficient for our analysis of polarization; hence, the vector represents the polarization state in terms of the amplitude ratio and retardance between the Cartesian electric field components $E_x$ and $E_y$. The aim of the system is to obtain a specified $\boldsymbol{J}_{out}$ in the presence of an unknown $\boldsymbol{J}_{sample}$. As we have control of $\boldsymbol{J}_{mod}$, we can obtain output measurements using different input polarizations in order to determine the unknown $\boldsymbol{J}_{sample}$. The Jones matrix of an arbitrary birefringent material could be simplified as [34]:

$$J_{sample} = \begin{bmatrix} \cos^2\theta + e^{i\phi}\sin^2\theta & (1-e^{i\phi})e^{-i\eta}\cos\theta\sin\theta \\ (1-e^{i\phi})e^{i\eta}\cos\theta\sin\theta & \sin^2\theta + e^{i\phi}\cos^2\theta \end{bmatrix} \quad (2)$$

where $\theta$ is the fast axis orientation with respect to the $x$-axis, $\phi$ is the retardance between $E_x$ and $E_y$, and $\eta$ is the circularity. Therefore, a minimum of three measurements needs to be taken to obtain the matrix. If $\eta = 0$, e.g. for linear retarders, the unknown variables are reduced to two. Further, if we have prior knowledge of the magnitude of the retardance of the perturbation, a single measurement is sufficient. Hence, in situations where we have different prior knowledge of the perturbation, we could apply different settings of $J_{mod}$ and measure the corresponding $J_{out}$. The Jones matrix of the sample could be calculated according to Eq. (1). After the $J_{sample}$ is obtained, we could calculate the $J_{mod}$ required for each desired output polarization state $J_{out}$. The PSG would then be used to generate the corresponding $J_{mod}$.

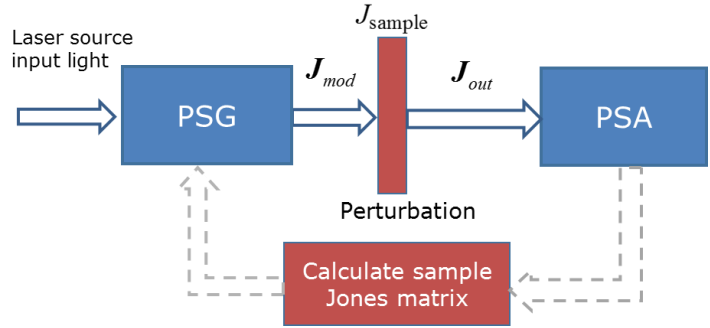

Fig. 1. Principle of the APC system operation. Blue solid line: beam propagation direction; grey dashed line: calculation path

## 2.2 Experimental set-up

A schematic of the APC system is shown in Fig. 2. This set-up is an enhanced version of that used in Ref [35]. The PSG used a dual-pass configuration using the SLM, which was able to generate any spatially variant polarization state. The PSA was able to calculate the polarization state of the output light by rotating the QWP to four different angles and recording the intensity profile of the beam. The SLM was pre-calibrated for retardance modulation [36] and the whole system was then calibrated by mapping the voltage applied on the SLM to the output polarization state measured by PSA. The perturbing sample was placed between the PSG and PSA. The sample was conjugated using 4f systems to the SLM in the PSG and to the CMOS camera in the PSA.

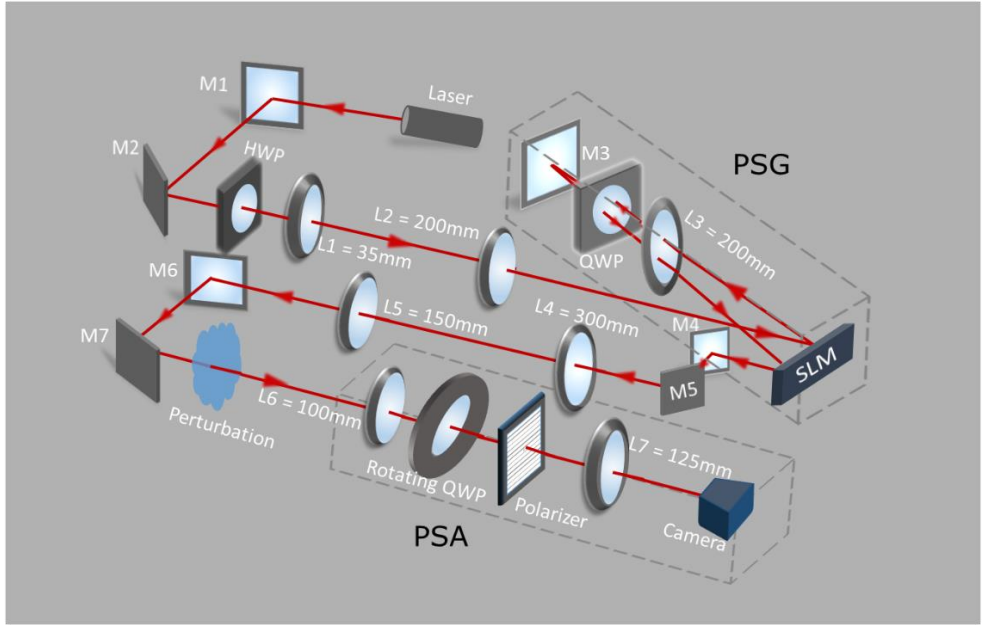

Fig. 2. Schematic of the experimental set-up of the APC system. The PSG and the PSA are shown within the dashed boxes. M1 to M7: mirrors. HWP: half wave plate with fast axis at 22.5°. L1 to L7: lens. SLM: spatial light modulator. QWP: quarter wave plate with fast axis at 22.5°. The fast axis of the polarizer was fixed to be horizontal and the orientations of the rotating QWP were set sequentially at 15.1°, 51.7°, 128.3°, and 164.9° when measuring the polarization state.

In the PSG, each pixel on the SLM could be regarded as a phase retarder with fixed horizontal fast axis. Its Jones matrix can be written as:

$$J_{SLM} = \begin{bmatrix} e^{i\frac{\delta}{2}} & 0 \\ 0 & e^{-i\frac{\delta}{2}} \end{bmatrix} \quad (3)$$

where $\delta$ is the retardance. Both of the HWP and QWP were rotated so that their extraordinary axes were at 22.5° to the x-axis. The Jones vector of the output light could be represented as:

$$\boldsymbol{J}_{out} = J_{SLM2}(\delta_2) J_{QWP}(-22.5°) J_{mirror} J_{QWP}(22.5°) J_{SLM1}(\delta_1) J_{HWP}(22.5°) \boldsymbol{J}_{in} \quad (4)$$

where $J_{mirror}$ represents the Jones matrix of the mirror, same for the other components; $\delta_1$ is the retardance introduced by the first pass of the SLM, $\delta_2$ is the retardance from the second pass and $\boldsymbol{J}_{in} = \begin{bmatrix} 0 & 1 \end{bmatrix}^T$ represents y polarized input light. The fast axis angle in the calculation of the output Jones vector is set along the beam propagation direction, therefore, in the PSG, after light is reflected by M3, the fast axis orientation of the QWP is used as -22.5°. We can then substitute this into all the Jones matrices to obtain the simplified output vector:

$$\boldsymbol{J}_{out} = \begin{bmatrix} e^{i\frac{-\delta_1+\delta_2}{2}} - e^{i\frac{\delta_1+\delta_2}{2}} \\ e^{i\frac{-\delta_1-\delta_2}{2}} + e^{i\frac{\delta_1-\delta_2}{2}} \end{bmatrix} = 2 \begin{bmatrix} \sin\frac{\delta_1}{2} e^{i\frac{\pi-\delta_2}{2}} \\ \cos\frac{\delta_1}{2} e^{i\frac{-\delta_2}{2}} \end{bmatrix} \quad (5)$$

We see that the amplitude ratio between $E_x$ and $E_y$ is determined by $\delta_1$, the first SLM pass retardance, and the phase shift between $E_x$ and $E_y$ is determined by the second pass retardance $\delta_2$. Therefore, assuming $\delta_1$ and $\delta_2$ can take any value between 0 and $2\pi$ radians, the polarization of the output light can be manipulated to any state by using the dual-pass SLM system.

Using the PSA, the polarization could be determined by a sequence of four measurements. The camera recorded the intensity of the output light for four different orientations of the QWP. The Stokes parameters were calculated as:

$$S_{out} = A^{-1}I \quad (6)$$

where $A$ is a 4×4 matrix. Each row in $A$ is the first row of the Müller multiplication of the polarizer and the QWP $M_pM_{QWP}(\theta)$, where $M_p$ is the Müller matrix of the linear polarizer, whose fast axis was fixed to be horizontal; $M_{QWP}(\theta)$ is the Müller matrix of the QWP, with the fast axis orientation of $\theta$. Matrix $A$ could then be determined when substituting $\theta = 15.1°$, $51.7°$, $128.3°$, and $164.9°$ respectively [37, 38]. The intensities obtained at each fast axis orientation $\theta$ were recorded in sequence and form a 4×1 matrix $I$. Hence, the output Stokes parameters can be calculated using Eq. (6). We refer to this as one PSA measurement.

As we are concerned here with restoring the desired polarization state through birefringence transformations, we make the implicit assumption that we are using fully polarized light. We can thus convert the Stokes vectors to the corresponding normalized Jones vectors [39], as in Eq. (1), which describe only the polarized component of the beam. Due to the nature of Müller polarimeter, it is not possible to retrieve the overall phase of the beam, although this could be obtained using alternative phase measurements.

## 3. Results and discussion

This APC system can be controlled in different ways depending on what prior assumptions can be made about the specimen's birefringent properties. In order to illustrate this, we demonstrated four different kinds of samples, each of which required different numbers of measurements for correction.

### 3.1 Vortex half wave retarder

The first sample was a vortex half wave retarder (Thorlabs, WPV10L-780). This consists of a spatially variant half wave plate, whose retardance value is fixed, but whose fast axis is distributed as shown in Fig. 3(a). If we use horizontal polarized light for illumination, the output polarization state will still be linearly polarized but with a spatially varying orientation. Fig. 3(b) plots the polarization ellipse orientation of the output light as measured by the PSA. Note that for the PSG and the PSA, for calculations and control purposes we divided the image into 15×15 sub-regions, over which each of the individual PSA intensity measurements used was averaged before calculating the polarization state in that region. Correspondingly, the same voltage was applied within each sub-region on the SLM.

According to Eq. (2), if the retardance of the phase retarder is known to be $\pi$ radians, the Jones matrix of the vortex half wave retarder could be simplified as

$$J_{sample} = \begin{bmatrix} \cos^2\theta - \sin^2\theta & 2\cos\theta\sin\theta \\ 2\cos\theta\sin\theta & \sin^2\theta - \cos^2\theta \end{bmatrix} \quad (7)$$

Hence, only one measurement is needed to get the value of $\theta$ modulo $\pi$ radians. We set the PSG to generate horizontal polarized input light, e.g. $J_{mod} = \begin{bmatrix} 1 & 0 \end{bmatrix}^T$. The output Jones vector using this illumination is therefore:

$$J_{out} = J_{sample}J_{mod} = \begin{bmatrix} \cos^2\theta - \sin^2\theta & 2\cos\theta\sin\theta \\ 2\cos\theta\sin\theta & \sin^2\theta - \cos^2\theta \end{bmatrix}\begin{bmatrix} 1 \\ 0 \end{bmatrix} = \begin{bmatrix} \cos^2\theta - \sin^2\theta \\ 2\cos\theta\sin\theta \end{bmatrix} \quad (8)$$

By using the PSA and converting the measured Stokes parameters to Jones vectors, $\theta$ can be simply calculated. After the sample Jones matrix was determined, given the desired output polarization state, expressed as the Jones vector $J_{d\_out}$, $J_{mod}$ can be worked out using:

$$J_{mod} = J_{sample}^{-1} J_{d\_out} \qquad (9)$$

As we wanted to restore the light to horizontal polarization after the sample disturbance, we calculated the $J_{mod}$ and applied the corresponding voltage on the SLM to create the necessary input polarization. Fig. 3(c) shows the polarization orientation after the APC correction. It is seen that the polarization orientation after correction is near uniform and is close to zero at all points across the field. The root mean square (RMS) variations of $S_1$, $S_2$ and $S_3$ before and after correction are compared in Table 1.

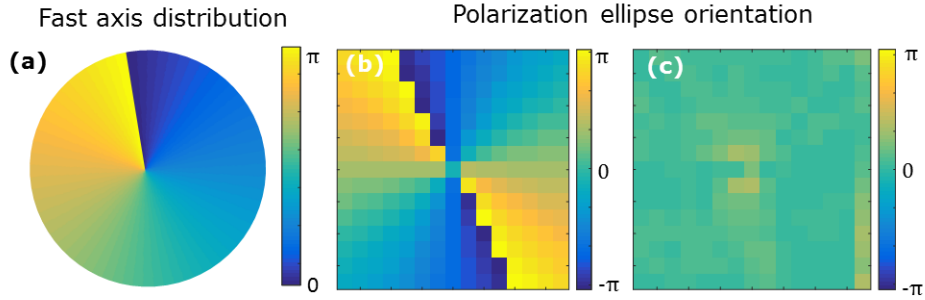

Fig. 3. (a) Fast axis distribution of the vortex half wave retarder. (b) Polarization ellipse orientation of the output light before APC correction. (c) Polarization ellipse orientation after APC correction.

Table 1. Comparison between Stokes parameters before and after correction

| Stokes Parameter | RMS before correction | RMS after correction |
| --- | --- | --- |
| $S_1$ | 1.0757 | 0.3630 |
| $S_2$ | 0.6060 | 0.1449 |
| $S_3$ | 0.1252 | 0.0751 |

As further illustrations of the correction process, Fig. 4 shows the output light intensity distribution on the camera viewed through two different orientation of the rotating QWP. Fig. 4(a) and (b) are the images taken at the rotating QWP fast-axis orientations of 15.1° and 164.9° before correction. The intensity pattern seen here is characteristic of what one should expect to see through the PSA following the vortex half wave retarder. Compared to that, the intensity distribution after correction are more uniform at the same QWP fast axis orientation as shown in Fig. 4(c) and (d), as the polarization state was restored to be horizontally polarized across the whole beam profile. The square grid patterns are due to pixel edge effects where the SLM has been divided into sub-regions.

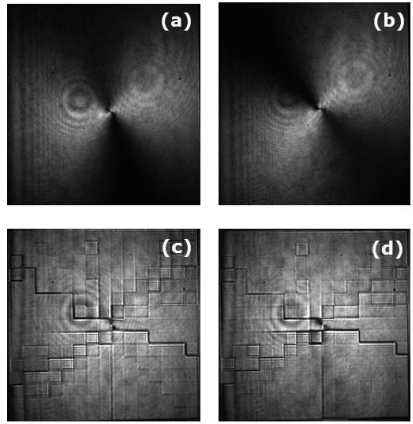

Fig. 4. Intensity distribution recorded from the camera at two different QWP fast axis orientations. (a) Before correction, QWP at 15.1°. (b) Before correction, QWP at 164.9°. (c) After correction, QWP at 15.1° (d) After correction, QWP at 164.9°. The intensity distribution is more uniform after correction (boundaries between the pixel regions can be seen, due to pixel edge effects in the SLM)

We also tested the APC system for reconstructing the output light to an arbitrarily chosen uniform polarization state. The target reconstructions for the output light were chosen to be horizontal, 45° linear and right circular polarized, respectively. Fig. 5 shows the Stokes parameters $S_1$, $S_2$ and $S_3$ of the output under different reconstructing situations, as retrieved by the PSA. It is seen that the APC system is able to manipulate the polarization state as expected. The discontinuity of the polarization state at the center region is because of the spatial sampling chosen for the sub-regions. This effect could be minimized by decreasing the sub-region size, but with reduced signal-to-noise ratio.

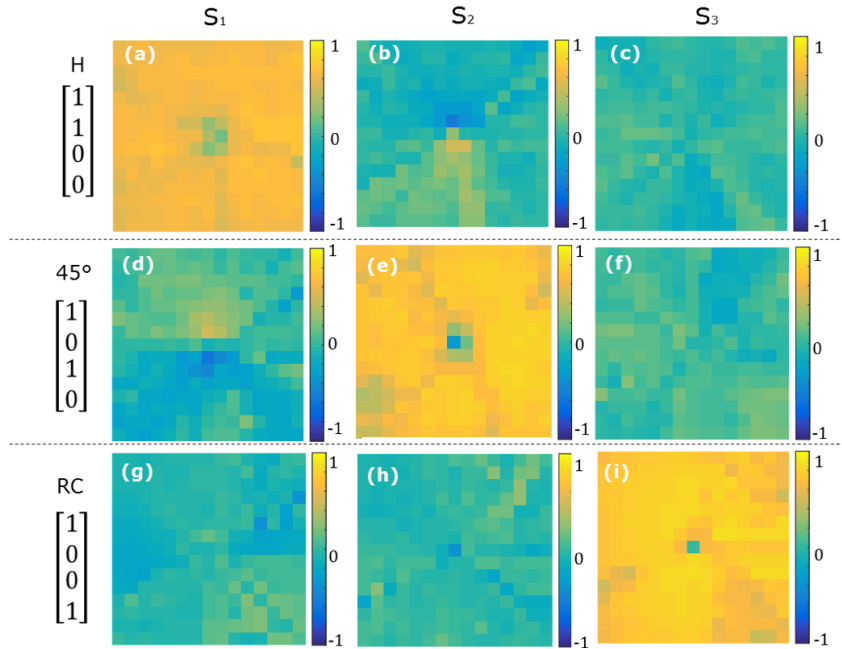

Fig. 5. Stokes parameters measured by the PSA under different reconstructing situations. (a) − (c): Stokes parameters $S_1$, $S_2$ and $S_3$ for horizontally polarized output (H). (d) – (f): Stokes parameters for 45° linearly polarized output. (g) – (i): Stokes parameters for right circular (RC) polarized output

## 3.2 Stressed plastic plate

Stressed polycarbonate is a biaxial birefringent material, whose in-plane and lateral refractive indices are different [40, 41]. The difference in the refractive index would introduce a retardance depending on the orientation of the optical axis of the plastic substrate. We tested a stressed plastic plate whose both retardance and fast axis orientation were unknown to demonstrate our system. The birefringent effect of the plastic plate sample viewed through crossed polarizers is shown in Fig. 6(a).

If the retardance and fast axis orientation are unknown, the Jones matrix of the sample can be represented as:

$$J_{sample} = \begin{bmatrix} \cos^2\theta + e^{i\phi}\sin^2\theta & (1-e^{i\phi})\cos\theta\sin\theta \\ (1-e^{i\phi})\cos\theta\sin\theta & \sin^2\theta + e^{i\phi}\cos^2\theta \end{bmatrix} \quad (10)$$

It is seen that the off-diagonal elements are equal. We can use this observation to extract the Jones matrix of the sample using two PSA measurements, which we chose to be taken with horizontally and vertically polarized illumination. The Jones vectors after using horizontal and vertical illumination are:

$$J_{out1} = J_{sample}J_{mod1} = \begin{bmatrix} A & B \\ B & C \end{bmatrix}\begin{bmatrix} 1 \\ 0 \end{bmatrix} = \begin{bmatrix} A \\ B \end{bmatrix} = \alpha_1 \begin{bmatrix} A_1 \\ B_1 \end{bmatrix} \quad (11)$$

$$J_{out2} = J_{sample}J_{mod2} = \begin{bmatrix} A & B \\ B & C \end{bmatrix}\begin{bmatrix} 0 \\ 1 \end{bmatrix} = \begin{bmatrix} B \\ C \end{bmatrix} = \alpha_2 \begin{bmatrix} B_2 \\ C_2 \end{bmatrix} \quad (12)$$

where $A$, $B$ and $C$ are all complex numbers in the sample Jones matrix. $A_1$, $B_1$ are the elements of the (normalized) Jones vector calculated from the conversion of Stokes parameters from the PSA of the first measurement, and $\alpha_1$ is the scaling factor to the original matrix element. Similarly, $\alpha_2$ is the equivalent factor for the second measurement. Since we know the two off-diagonal elements are the same, the relationship between the scaling factor $\alpha_1$ and $\alpha_2$ can be easily calculated and the normalized Jones matrix of the sample was then determined. Eq. (9) could then be used to determine the necessary input light field to produce the desired output.

Fig. 6(b) shows plots of the polarization ellipse of each sub-region before correction when using vertical polarized light to illuminate the plastic plate. The output polarization state is non-uniformly distributed across the field because of the uneven birefringence. Fig. 6(c) − (e) show the polarization ellipses under different reconstruction situations by using the APC system. It is seen that the system is able to reconstruct across most of the field the output polarization state, even with an unknown retardance and fast axis orientation of the perturbation. The mismatch of a few regions in Fig. 6(d) and (e) is because those sub-regions are areas of rapidly changing birefringence, which can be seen from the inset aperture of the illumination region from Fig. 6(a). Since the Jones matrix calculated for each sub-region is based on the polarization state measured by PSA, and the intensity was averaged for each sub-region when calculating the polarization state by using Eq. (6), there could still be mismatches in correction within each region.

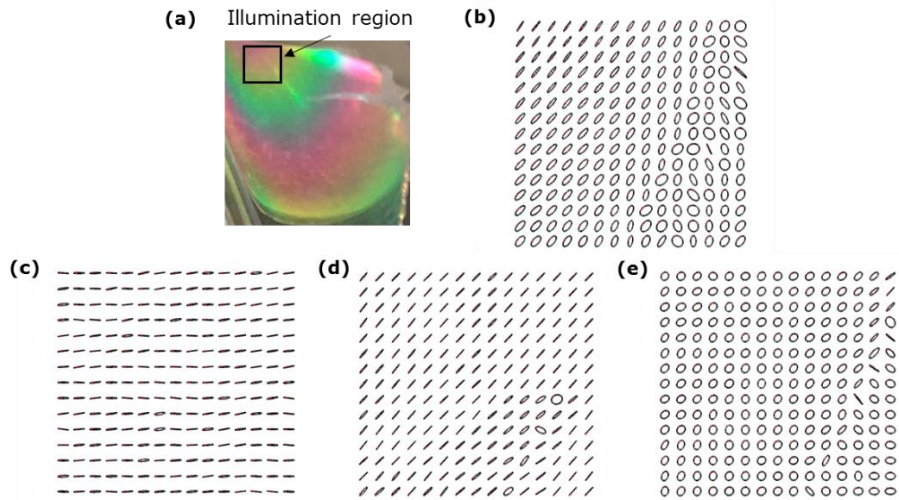

Fig. 6. (a) Birefringent effects of a stressed plastic plate illuminated by a lamp when viewed through cross polarizers with white light illumination. (b) Plotted polarization ellipse of the output before correction when using vertical polarized illumination. (c) Polarization ellipse under horizontal polarization state reconstruction (d) Polarization ellipse under 45° linear polarization state reconstruction. (e) Plotted polarization ellipse of the output light under right circular polarization state reconstruction.

*3.3 Liquid crystal device*

A liquid crystal (LC) device was used to show how the APC could adapt to changes in a dynamic system. This LC device was a 20 µm thick anti-parallel rubbed cell (Instec Inc). It had a 5×5 mm electrode region and was filled with E7 liquid crystal. We connected the liquid crystal device to a signal generator and placed it at the sample plane. The device was designed to have two different regions, which we refer to as the static region and the dynamic region. The birefringence in the static region, which was away from the electrode structure, did not vary with the external voltage, denoted $U$, while the dynamic region, in the vicinity of the electrodes, changed with a varying drive voltage. The schematic of the device is shown in Fig. 7(a). We tested the system's dynamic functionality by changing the voltage to the device and to see if APC could react to the different birefringence disturbance and reconstruct to the same polarization state.

Fig. 7(b) − (e) shows the measurements represented on a Poincaré Sphere when the sample was illuminated with a horizontally polarized light. Fig. 7(b) shows the output polarization state at different sub-regions (one point corresponds to one sub region) before correction when the drive voltage $U = 1.8$ volts. The output polarization state of the static and dynamic regions were distributed away from the desired horizontal polarization state because of the spatially varying birefringent perturbation introduced by the LC cell. After applying correction using the APC system, the polarization state after correction then became much more uniform and concentrated to the horizontal polarization state as shown in Fig. 7(c). Fig. 7(d) plots the output polarization state without APC compensation when the drive voltage was set to $U = 2.5$ volts. It can be seen that the polarization state of the reference region is similar to the equivalent measurements shown in part (b); due to the birefringence change in the dynamic region, the polarization state of that region had a different distribution. Despite the disturbance change, after APC correction the polarization of the output light still could be reconstructed to the uniform horizontal polarization as shown in Fig. 7(e). With this customized sample, we demonstrated that the APC is able to track dynamically changing perturbations.

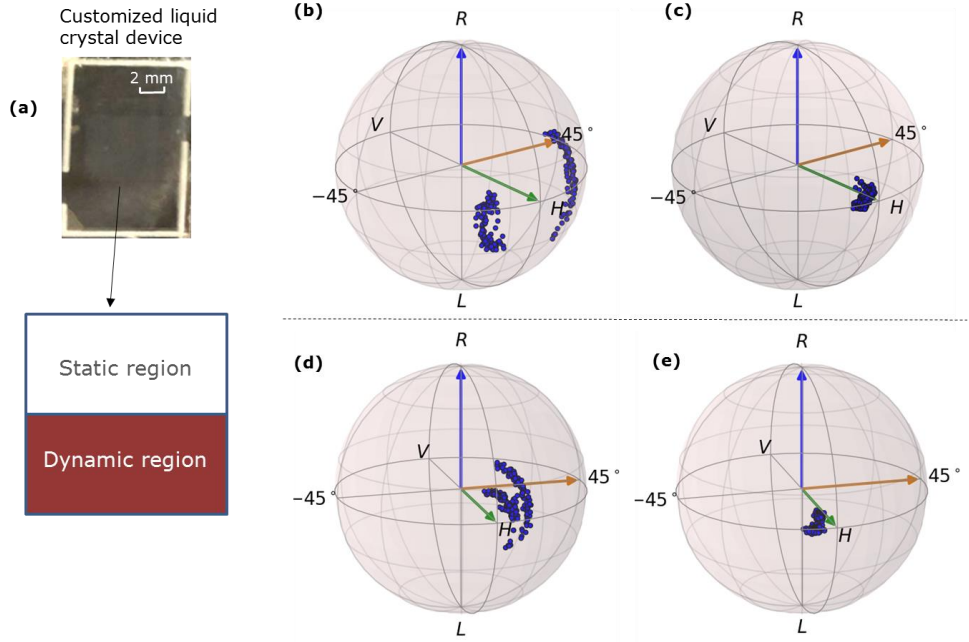

Fig. 7. (a) Schematic of the liquid crystal device, the birefringence of the dynamic region can be changed according to drive voltage while the static region does not. (b) Output light polarization state of each sub-region representing on Poincaré Sphere before correction, at $U = 1.8$ volts. (c) Polarization state distribution after correction at $U = 1.8$ volts. (d) Polarization state distribution before correction at $U = 2.5$ volts. (e) Polarization state distribution after correction at $U = 2.5$ volts.

## 3.4 Mouse brain tissue

The system was also demonstrated in a more general scenario, where no assumptions were made about the nature of the birefringent effects. When the disturbance is an arbitrary birefringent material showing zero attenuation, three measurements are needed to reconstruct the polarization state. We tested the operation of the APC system in this regime using a sample of mouse brain tissue [42, 43]. We used a thin tissue sample here to ensure minimal light loss and diattenuation introduced from the sample. The sample was a 60-µm thick fixed coronal brain slice from a Thy1-GFP line M mouse. A transmission microscope image of the tissue is provided in Fig. 8(a).

According to Eq. (2), to measure the sample Jones matrix, we need to take three independent measurements; we implemented this by controlling the PSG to generate horizontal, vertical and 45° linearly polarized illumination light, respectively. The output Jones vectors then can be represented as:

$$\boldsymbol{J}_{out1} = \boldsymbol{J}_{sample}\boldsymbol{J}_{mod1} = \begin{bmatrix} A & B \\ C & D \end{bmatrix}\begin{bmatrix} 1 \\ 0 \end{bmatrix} = \begin{bmatrix} A \\ B \end{bmatrix} = \alpha_1 \begin{bmatrix} A_1 \\ B_1 \end{bmatrix} \quad (13)$$

$$\boldsymbol{J}_{out2} = \boldsymbol{J}_{sample}\boldsymbol{J}_{mod2} = \begin{bmatrix} A & B \\ C & D \end{bmatrix}\begin{bmatrix} 0 \\ 1 \end{bmatrix} = \begin{bmatrix} C \\ D \end{bmatrix} = \alpha_2 \begin{bmatrix} C_2 \\ D_2 \end{bmatrix} \quad (14)$$

$$\boldsymbol{J}_{out3} = \boldsymbol{J}_{sample}\boldsymbol{J}_{mod3} = \begin{bmatrix} A & B \\ C & D \end{bmatrix}\begin{bmatrix} 1 \\ 1 \end{bmatrix} = \begin{bmatrix} A+B \\ C+D \end{bmatrix} = \alpha_3 \begin{bmatrix} \alpha_1(A_1+B_1) \\ \alpha_2(C_2+D_2) \end{bmatrix} \quad (15)$$

where $A$, $B$, $C$ and $D$ are complex elements of the sample's Jones matrix; $A_1$, $B_1$, $C_1$ and $D_1$ are normalized output Jones vector elements; $\alpha_1$, $\alpha_2$ and $\alpha_3$ are scaling factors between the matrix elements and normalized output Jones vectors. $A$ and $B$ can be first determined using the horizontal polarized illumination. The sample normalized Jones matrix can be determined if we find the relationship between $\alpha_1$ and $\alpha_2$ in Eq. (13) and (14). This can be worked out by having the additional measurement, Eq. (15).

A control measurement was performed using an empty region of the microscope slide adjacent to the brain tissue; this showed negligible birefringence effects, where the RMS of the variation the Stokes parameters were all less than 0.01. Fig. 8(b) and (c) show, as examples, the intensity images of the recorded by the camera when using horizontally polarized light illumination, with the QWP oriented to 128.3° before and after restoration to horizontal polarization, respectively. It can be seen from the uneven intensity distribution before compensation and the more uniform intensity after compensation that the brain tissue introduced a perturbation through birefringence. Fig. 8(d) − (f) give the Stokes parameters of $S_1$, $S_2$ and $S_3$ calculated from the PSA before compensation. The variance across the sample in $S_1$, $S_2$ and $S_3$ was then corrected by the APC after the restoration as shown in Fig. 8(g) − (i). These results demonstrate the functionality of this system using three measurements to compensate for an arbitrary birefringent sample. Considering that the brain tissue could also introduce some depolarization, we then measured the degree of polarization (DOP) before and after correction. It is seen that although some depolarization is present, the APC system could correct the polarized component without significant changes to the DOP, from a mean value of 0.82 to 0.79. The statistical distribution of Stokes parameters and DOP comparison before and after correction are shown in Fig. 9. The variation of the Stokes parameters across the field of view was clearly reduced by the APC system.

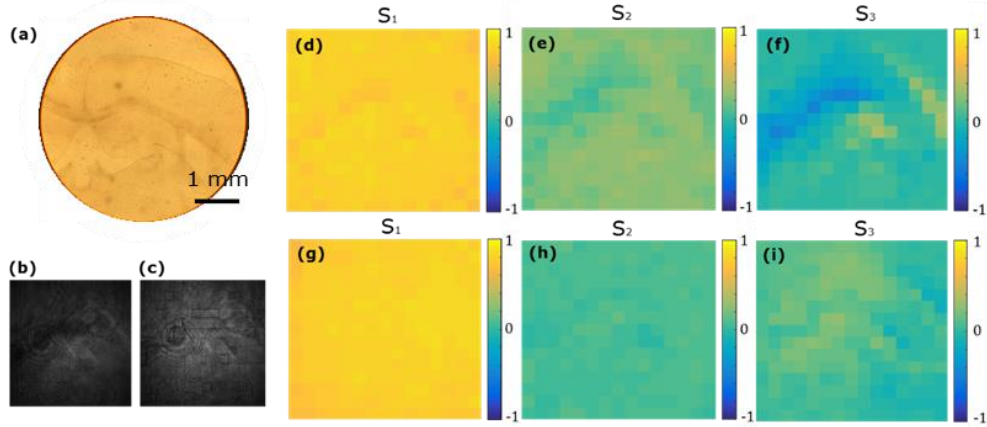

Fig. 8. (a) Microscope image of the mouse brain tissue sample. (b) Intensity image recorded by the camera of the brain tissue sample before correction at the rotating QWP of 128.3°. (c) Intensity image of the sample after correction. (d) − (f) $S_1$, $S_2$ and $S_3$ of the polarization state after the brain tissue when using horizontal polarized illumination before correction. (g) − (i) $S_1$, $S_2$ and $S_3$ of the polarization state following correction.

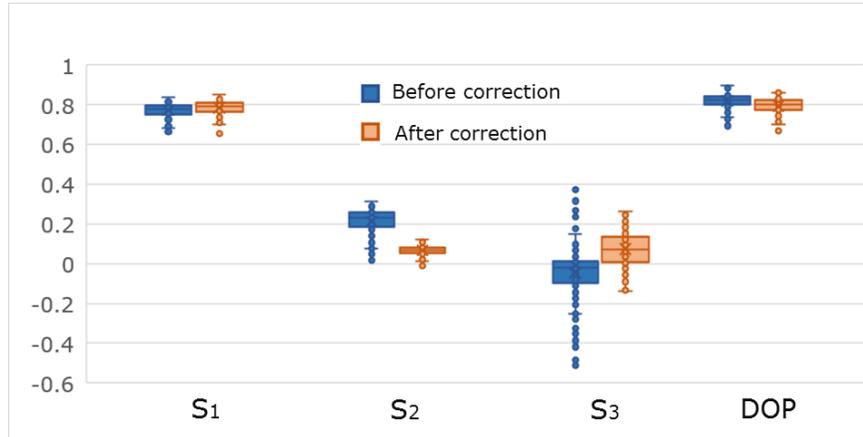

Fig. 9. Stokes Parameters, $S_1$, $S_2$, $S_3$ and DOP comparison before and after correction. Each data point represents the parameter of a sub-region. The bars show mean and standard deviation.

## 4. Discussion

The demonstrations present here illustrate possibilities for implementation of APC in different practical scenarios. In principle, for non-diattenuating specimens, there are three unknowns in the Jones matrix, and hence three independent measurements of the output Jones vector are required. If unknown diattenuation were also present, then more measurements would be required to determine the optical properties of the specimen. The number of measurements required in practice depends upon the assumptions one can make about the specimen, where knowledge about, for example, the retardance or orientation of the birefringence can be used. Using such prior knowledge would help increase the speed of operation of the APC system. The current implementation uses sequential measurements with four positions of the rotating waveplate in the PSA for each Stokes vector (or equivalently Jones vector) measurement. If speed was of the essence, then it would be possible to employ other forms of imaging polarimeter for single shot measurement [24, 25].

The demonstrations here have concerned purely restoration of the polarization profile of the beam to its desired state. However, as discussed in the introduction, phase aberrations also play an important role in system performance. The methods described in this paper could be combined with conventional phase adaptive optics methods to provide further benefits in system correction.

## 5. Conclusion

Through implementation of the APC, we have shown how external complex perturbations of the polarization profile across a beam can be corrected using appropriate feedback from the PSA to the PSG. The system presented here could form the basis of an online polarization compensation system in optical systems that are affected by spatially variant polarization errors. Such an approach would have benefit in applications ranging from quantum optics to microscope imaging.

**Funding**

European Research Council (ERC) under the Horizon 2020 research and innovation program (AdOMiS, grant agreement No. 695140). China Scholarships Council.

**Acknowledgement**

The authors would like to thank John Sandford O'Neill for his help with the liquid crystal device sample preparation.

**Disclosure**

The authors declare no conflicts of interest.